\documentclass[11pt,a4paper]{article}%\usepackage[makeroom]{cancel}
\usepackage{bm}
\usepackage{amsmath}
\usepackage{amsfonts}
\usepackage{amssymb}
\usepackage{graphicx}
\usepackage{authblk}
\usepackage{cite}
\usepackage[left=2.4cm,top=2.4cm,right=2.4cm,bottom=2.4cm,nohead]{geometry}
\DeclareGraphicsExtensions{.png,.jpg,.pdf}
\usepackage{epstopdf}
\usepackage{float}
\usepackage{fouriernc} 
\usepackage{enumitem}
\usepackage[utf8]{inputenc}
\usepackage[english]{babel}
\usepackage{mathtools}
\usepackage{amscd}
\usepackage{mathrsfs}
\usepackage{amsthm}
\usepackage{siunitx}%For SI units
\usepackage{caption}
\usepackage{subcaption}
\setlist{nolistsep,leftmargin=*}

\usepackage{mhchem} 
\usepackage{cleveref} 
\usepackage{xcolor}
\usepackage{soul}
\usepackage{mdframed}
\usepackage{framed}
\usepackage{nicefrac}
\usepackage{longtable}
\DeclareMathAlphabet{\mathpzc}{OT1}{pzc}{m}{it}

\title{Theory of expansion and compression of polymeric materials}

\makeatletter

\renewcommand\AB@authnote[1]{\textsuperscript{\normalfont#1}}

\makeatother

\author[1]{P.M. Biesheuvel,}
\author[2]{H. Fan,} 
\author[2]{M. Elimelech}

\affil[1]{Wetsus, European Centre of Excellence for Sustainable Water Technology,  Leeuwarden, The~Netherlands.}
\affil[2]{Department of Chemical and Environmental Engineering, Yale University, New Haven, USA.}

\date{} 

\newcommand{\s}[1]{\mathrm{_{#1}}}
\begin{document}

\maketitle

\begin{abstract}

We extend classical Flory-Rehner theory for the expansion and compression of porous materials such as cross-linked polymer networks. The theory includes volume exclusion, affinity with the solvent, and finite stretching of the polymer chains. We also modify this equilibrium theory --that applies to equal expansion of a material in all directions-- to the situation that a material can only expand in a single direction, as is the case when a thin layer is tightly bound to a support structure. We extend this equilibrium model to the case that a pressure is applied across such a thin layer of the polymer material, for instance a membrane, and liquid flows across this layer. The theory describes how in the direction of liquid flow the membrane is increasingly compacted (becomes less porous), and the more so at higher applied pressures. We provide results of example calculations for a thick membrane with significant changes in compaction across its thickness, and a thin membrane for which compaction due to flow is minor. In the last section we model the dynamics of the change of size of a porous material in time after a step change in the solvent-polymer attraction parameter.

\end{abstract} 

\section{Introduction}

Elastic porous materials, for instance made of cross-linked polymer chains, play an important role in many applications including tissue engineering and membranes for water desalination~\cite{Horkay_2007,Davenport_2020}. In this work we focus on polymeric materials that form cross-linked networks with open spaces, i.e., pores or free volumes, in between the polymer chains. The degree of swelling of the material can be expressed as a porosity, which is the volume fraction that is open and accessible for solvent and solutes such as ions. This porosity depends on many properties of the polymer network
such as cross-linking degree, elasticity, and polymer-solvent interaction. When the polymer is charged, electrostatic (Donnan) effects also play a role, but these effects are not discussed in the present work. 

The classical treatment is based on the Flory-Rehner theory~\cite{Flory_Rehner_1943}, which %heuristically 
combines lattice-based Flory-Huggins polymer theory that describes the interaction between polymer segments, with a linear theory for chain elasticity that assumes chain stretching is low or moderate. The  assumption of low to moderate stretching may not be valid in practice, resulting in an overestimate of the volume of a material, while the lattice-based description of polymer interaction is difficult to combine with theory for hydrodynamics and fluid flow through the porous material. The lattice-based description of interaction also underestimates the importance of volumetric interactions between polymers, and thus it will overestimate of the density of the material, i.e., it is not an accurate equation of state for polymer solutions. %The derivation in ref.~\cite{Flory_1943} is difficult to follow, but the derivation in ref..  

In the present work we set up a theory for polymer expansion (swelling) and compression (compaction) that includes volume exclusion by the polymer chains based on an accurate equation of state, a solvent-polymer interaction energy, and non-linear theory of chain elasticity that includes finite stretching of the cross-linked polymer network. We first describe how to improve the Flory-Rehner theory to describe compression and expansion in three dimensions (directions) of a cross-linked polymer material, such as a hydrogel particle, and next describe how this theory changes when we consider a thin layer that can only expand in one direction. Finally we extend this equilibrium model to the situation that solvent flows across a thin layer of such a porous material when a pressure difference is applied. We include expressions for fluid-polymer friction that depend on the density of the material. One type of such a porous layer are membranes, which are are applied in many separation processes in industry and water treatment, both for organic solvents and aqueous solutions. Calculations show how in the direction of solvent flow (towards lower pressures), the material becomes more dense, i.e., the porosity goes down. By how much the porosity decreases depends on properties of the membrane such as permeability and thickness, and on the pressure applied.

\section{Theory for equilibrium swelling of cross-linked polymer networks}

In this section we provide theory for the equilibrium swelling degree of a porous polymer material submerged in a volume of solvent, and this solvent will also fill all pore space in the material. Such a cross-linked polymer material % (in this work also called membrane, network, or polymer film) 
has a porosity, \textit{p}, or, vice-versa, a polymer volume fraction, $\phi$. These two volume fractions add up to one, thus $p+\phi=1$. The volume fraction of polymer, $\phi$, is the fraction of the total volume of a material that is occupied by polymer chains, and $p$ is the volume fraction filled %The open volume, i.e., the pores, or porosity, is filled 
with solvent (and if present, solutes).\footnote{In this paper the terms solvent, liquid, and fluid, all have the same meaning.} When the material is compressed, porosity goes down, and when it swells, porosity goes up. Because the total amount of polymer in a certain particle is fixed, a change in porosity of the particle submerged in liquid implies that the volume of the particle will change while for some time liquid flows into or out of the particle, because the pore phase (porosity) is always filled with solvent.

The polymer network consists of nodes (also called cross-linkages, or branching points), with polymer chains (also called sub-chains, strings, or strands) in between the nodes. These chains oppose being stretched, i.e., a force is needed to stretch them, and progressively more force is needed to stretch them further. We assume the material only contains one type of chain, of one length, so --when a material is free to expand in all three directions-- at the same macroscopic location all chains are stretched equally. 
To describe polymer expansion and compression, we define several parameters.  
First of all, there are \textit{N} polymer chains per unit total volume, each with contour length $\ell$, which is the length, or distance, along the contour, or backbone, of the polymer chain from one node to the next. Thus $N\!\cdot\!\ell$ is a `polymer length density'. On both its ends, each chain is terminated by nodes, where a chain is connected to two %or three 
other chains. %in this way forming a cross-linked network. 
In part of the theory the chains are described as strings of touching beads, each bead with size $\sigma\s{b}$ and volume $v\s{b}=\pi/6\cdot \sigma_\text{b}^3$. This approach leads to an accurate equation-of-state for polymer solutions, brushes, and gels, that is in alignment with results of MD simulations~\cite{Biesheuvel_2008,Spruijt_2014,Spruijt_2015,Biesheuvel_Dykstra_2020}. With a polymer fraction $\phi$, the concentration of these beads is $c\s{b}=\phi/v\s{b}$. The various parameter just introduced are related by $N \cdot\ell = c\s{b} \cdot \sigma\s{b}$.

We first consider the case that a porous material is free to expand and contract in all directions, and then any relative volumetric change leads to a relative change in any distance measure that is one-third the volumetric change, i.e.,
\begin{equation}
\frac{1}{x} \text{d}x = \frac{1}{3} \frac{1}{V} \text{d}V
\label{eq_defintion_expansion_volume}
\end{equation}
where \textit{x} can be a macroscopic distance between any two points in the gel, or a microscopic distance between two nodes that are on either end of a chain. This is similar to Eq.~(2) in ref.~\cite{Flory_Rehner_1943} where the extension of chains and the polymer volume fraction are related by the same proportionality. We define the chain stretching degree, $\widetilde{x}$, as the actual distance between chain end-points (nodes), $x$, divided by the contour length, $\ell$, i.e., $\widetilde{x}=x/\ell$, and we then arrive for a material that can expand in three directions at\footnote{In the present theory it is assumed that all chains are equally long (along their backbone), and that there are no topological constrictions (entanglements) in the network, but all chains can expand or contract freely without such restrictions.} 
\begin{equation}
\left(\frac{\widetilde{x}}{\widetilde{x}_0}\right)^{\text{~}3} =\frac{\phi\s{0} }{ \phi}  \hspace{5mm},\hspace{5mm}  {\widetilde{x}}^{\text{~}3} =\frac{\phi\s{min} }{ \phi}  
\label{eq_x_to_eta_min}
\end{equation} 
where $\widetilde{x}_0$ is a reference stretching degree, and $\phi_0$ the polymer density at that reference condition. One option is to choose $\widetilde{x}_0=1$ and then $\phi_0$ is the minimum polymer density, $\phi\s{min}$, that we reach when all chains are fully stretched, i.e., for that condition $x=\ell$, thus $\widetilde{x}\!=\! 1$.

To calculate the density of a porous cross-linked polymer material, we must analyze several pressures, and at mechanical equilibrium they add up to zero. These pressures are due to: 1. volumetric interactions between polymer segments, $\Pi\s{exc}$, where index `exc' refers to `excess' or `excluded volume'; 2. a polymer-solvent interaction, $\Pi\s{aff}$; and 3. a pressure exerted on the network because of the elasticity of the network, $\Pi\s{el}$.

In the absence of solutes (such as ions) in the solvent, the pressure due to volumetric interactions between the polymer chains is~\cite{Biesheuvel_2008, Spruijt_2014, Spruijt_2015, Biesheuvel_Dykstra_2020}
\begin{equation}
\Pi\s{exc} \cdot \frac{v\s{b}}{k\s{B}T} = \frac{\phi^2\left(3-\phi^2\right)}{\left(1-\phi\right)^3}  \sim 3 \phi^2 + 9 \phi^3 + 17 \phi^4 + \dots
\label{eq_bmcsl_sb}
\end{equation}
where $k\s{B} T$ is the thermal energy, which at room temperature is $k\s{B}T \! = \!  4.11\cdot 10^{-21}$~J. Eq.~\eqref{eq_bmcsl_sb} is based on the Carnahan-Starling equation-of-state for hard sphere mixtures, extended to the situation that the spheres in such a mixture form long or connected chains~\cite{Biesheuvel_2008,Spruijt_2014,Spruijt_2015,Biesheuvel_Dykstra_2020}. The polymer also has chemical interactions with the solvent, i.e., an affinity, and when water is the solvent this is often described by the terminology of hydrophilicity and hydrophobicity. Its contribution to the total pressure is given by~\cite{Biesheuvel_2008, Spruijt_2014, Spruijt_2015, Biesheuvel_Dykstra_2020}
\begin{equation}
\Pi\s{aff} \cdot \frac{v\s{b}}{k\s{B}T} = - \chi \phi^2
\label{eq_Pi_affinity}
\end{equation} 
where we make use of the index `aff' for affinity. The interaction parameter $\chi$ is positive when solvent-polymer interaction is unfavourable. Then the two materials prefer being demixed, which opposes the polymer network from swelling. Vice-versa, a negative value of $\chi$ implies that solvent and polymer like to be mixed on the molecular scale, which then results in a force that swells the particle. 

At equilibrium (i.e., without an external force and in the absence of solvent flowing through the material), these two forces are exactly compensated by a contractive elastic pressure, which we discuss next. This pressure originates from the fact that a polymer chain consists of many segments that are more or less freely jointed. When they are freely jointed, the angle/orientation of each segment is unrelated to that of its neighbours, leading a configuration similar to a random walk in three dimensions. These freely-jointed segments are called Kuhn segments. With one chain end fixed at a certain position, $\mathcal{O}$, and the other chain end located along a line $\mathscr{L}$ that starts in $\mathcal{O}$, the number of configurations to find the other end at a certain distance $x$ away from $\mathcal{O}$ along this line $\mathscr{L}$ decreases with \textit{x}, making these configuration that have a higher chain extension less likely. This translates into a contractive force between the two chain ends (between the nodes) which pulls the entire network closer. 
As we discuss next, this force increases with distance \textit{x}, and diverges when \textit{x} approaches the contour length, $\ell$.

The Langevin equation describes the force, $\widetilde{f}$, by which the two chain ends pull on each other as function of stretching degree, $\widetilde{x}$, and is given by~\cite{Kuhn_1946,Treloar_1975,Doi_1996,Biesheuvel_2004,Jedynak_2015} 
\begin{equation}
\widetilde{x}= \frac{1}{\tanh \left( k \, \widetilde{f} \right)}- \frac{1}{k \, \widetilde{f}}
\label{eq_Langevin}
\end{equation}
where $k$ is the Kuhn length, i.e., the length of a Kuhn segment. A shorter Kuhn length implies that we have more Kuhn segments in a given chain, and thus the polymer material is more difficult to stretch. The series expansion around $\widetilde{x} =0$ is given by \cite{Kuhn_1946,Treloar_1975}
\begin{equation}
 {k \, \widetilde{f}} =  \frac{k \, f}{k\s{B}T} =   3 \, \widetilde{x} + \frac{9}{5} \, \widetilde{x}^3  +  \frac{297}{175} \, \widetilde{x}^5 + \frac{1539}{875} \, \widetilde{x}^7 +\mathcal{O}\left(\widetilde{x}^9 \right) 
\label{eq_Kuhn}
\end{equation}
which for low to moderate $\widetilde{x}$ simplifies to 
\begin{equation}
k \widetilde{f} = 3  \cdot \frac{x}{\ell}  = 3  \cdot  \widetilde{x} \, .
\label{eq_Hooke}
\end{equation}
There are no exact solutions to the inversion of Eq.~\eqref{eq_Langevin}, expressing force explicitly as function of stretching degree, but the ``Cohen exact Padé approximation [3/2]'' is close, and has the prefactors up to $\widetilde{x}^5$ correct, resulting in~\cite{Jedynak_2015}
\begin{equation}
 {k \, \widetilde{f}} \sim \widetilde{x} \, \frac{3 - \frac{36}{35} \widetilde{x}^2 }{1- \frac{33}{35} \widetilde{x}^2 } \sim 3 \, \widetilde{x} + \frac{9}{5} \, \widetilde{x}^3  +  \frac{297}{175} \, \widetilde{x}^5 + \mathcal{O}\left(\widetilde{x}^7 \right) 
\label{eq_Pade}
\end{equation}
with the prefactor for $\widetilde{x}^7$ an underestimate of the exact result by 10\%, and by 20\% for $\widetilde{x}^9$. The ``rounded'' version of this equation is
\begin{equation}
 {k \, \widetilde{f}} \sim \widetilde{x} \, \frac{3 - \widetilde{x}^2 }{1-  \widetilde{x}^2 }= 3 \, \widetilde{x} + 2 \,  \widetilde{x}^3  +  2 \, \widetilde{x}^5 + \mathcal{O}\left(\widetilde{x}^7 \right) 
\label{eq_Pade_rounded}
\end{equation}
which overestimates the prefactors from $\widetilde{x}^3$ onward, by 10\% for $\widetilde{x}^3$, 18\% for  $\widetilde{x}^5$, and 25\% for $\widetilde{x}^7$, so also this approximation is not that far off.

The limit of Eq.~\eqref{eq_Langevin} for $\widetilde{x} \rightarrow 1$ is~\cite{Naji_2003}
\begin{equation}
k \widetilde{f} = \frac{1}{1-\widetilde{x}}
\label{eq_Langevin_high_stretching_limit}
\end{equation}
and this equation is very close to Eq.~\eqref{eq_Langevin} for stretching degrees beyond 50\%. 

The elastic energy of a single chain is 
\begin{equation}
U\s{el} =   \int_0^x f \text{d} x
\label{eq_definition_elastic_energy}
\end{equation}
which for the low to moderate stretching limit, Eq.~\eqref{eq_Hooke}, results in
\begin{equation}
U\s{el} = k\s{B}T \cdot \frac{\ell}{k} \cdot \frac{3}{2} \cdot \widetilde{x}^2 \, .
\label{eq_Hookean_energy}
\end{equation}
When we also include higher order terms, we have~\cite{Kuhn_1946,Treloar_1975,Doi_1996,Biesheuvel_2004} 
\begin{equation}
{U\s{el}} = {k\s{B}T} \cdot \frac{\ell}{k} \cdot \sum_i a_i \cdot \widetilde{x}^{2 i}
\label{eq_elastic_higher_order} 
\end{equation}
where index \textit{i} runs from 1 to $\infty$, and where the prefactors up to $\widetilde{x}^8$ are $a_1=3/2$, $a_2=9/20$, $a_3=99/350$, and $a_4=1539/7000$. 

The limit for high stretching ($\widetilde{x}>0.5$), Eq.~\eqref{eq_Langevin_high_stretching_limit}, can be integrated, which leads to~\cite{Naji_2003, Biesheuvel_2004}
\begin{equation}
U\s{el}=- \, k\s{B}T\cdot \frac{\ell}{k} \cdot \left(\ln\left(1-\widetilde{x} \right)+f_1\right)
\end{equation}
where the factor $f_1$ is given by $f_1 \sim 0.307$.

To go from an energy per chain to an elastic pressure, $\Pi\s{el}$, we must evaluate
\begin{equation}
\Pi\s{el} = - \frac{\partial \, V \! N U\s{el}  }{\partial \, V} = - V \! N \,   \frac{\partial U\s{el}}{\partial V}
\label{eq_osm_pr_definition}
\end{equation}
where $V$ is the volume of a porous polymer particle. We can treat the term $V \! N$ as a constant because the differentiation is performed under the constraint of a constant amount of polymer (in the changing volume), and the amount of polymer is proportional to $V \! N$. The elastic pressure, $\Pi\s{el}$, is a force that pulls the material closer together, i.e., it is a contractive force, and thus it is a negative quantity. 

We implement Eq.~\eqref{eq_defintion_expansion_volume} in~Eq.~\eqref{eq_osm_pr_definition} and arrive at
\begin{equation}
\Pi\s{el}  = - N \cdot \frac{x}{3} \cdot \frac{\text{d} U\s{el}}{\text{d} x}  = - N \cdot \frac{\widetilde{x}}{3} \cdot \frac{\text{d} U\s{el}}{\text{d} \widetilde{x}} =  - N \cdot \frac{x}{3} \cdot f = - \frac{\sigma\s{b}}{k} \cdot \frac{\widetilde{x}}{3} \cdot \frac{\phi}{v\s{b}} \cdot k\widetilde{f} \cdot k\s{B}T
\label{eq_osm_pr_definition_2}
\end{equation}
where we also included Eq.~\eqref{eq_definition_elastic_energy}. Thus the elastic pressure is proportional to the contractive force per chain, \textit{f}, to the number of chains per unit volume, \textit{N}, and to the stretching degree, \textit{x}. 

For low levels of stretching, we can implement Eqs.~\eqref{eq_x_to_eta_min}~and~\eqref{eq_Hooke} in Eq.~\eqref{eq_osm_pr_definition_2}, which leads to
\begin{equation}
\frac{\Pi\s{el}}{k\s{B}T} = - N \cdot \frac{\ell}{k} \cdot \widetilde{x}^2 = - N \cdot \frac{\ell}{k} \cdot \widetilde{x}_0^2 \cdot \left(\frac{\phi_0 }{\phi}\right)^{2/3}  = - \frac{1}{v\s{b}} \cdot \frac{\sigma\s{b}}{k} \cdot \widetilde{x}_0^2 \cdot \phi\s{0}^{2/3} \cdot \phi^{1/3} 
\label{eq_osm_pr_hookean}
\end{equation}
where we made use of $N  \ell = \sigma\s{b}  c\s{b}$ and $ c\s{b} = \phi / v\s{b}$. Eq.~(15) in ref.~\cite{Katchalsky_1954} gives $\Pi\s{el}$ as function of particle volume to the power --1/3, in line with Eq.~\eqref{eq_osm_pr_hookean} where $\Pi\s{el}$ depends on $\phi$ to the power 1/3. In a footnote on p.~23 in ref.~\cite{Katchalsky_1954}, we find that the next terms scale with $\phi$ to the power --1/3, --1, etc, which is also what we find.

In general, at each position in a porous material, we have the force balance~\cite{Van_der_Sman_2015}
\begin{equation}
P^\text{c} = \Pi\s{exc} + \Pi\s{aff} + \Pi\s{el} 
\label{eq_polymer_force_balance}
\end{equation}
where $P^\text{c}$ is the compression pressure. When the material is changing its shape or volume in time (it is shrinking or swelling), or when there are flows through the material, or when it is pushed against a surface, in all these cases $P^\text{c}$ will not be zero. But when the shape of the material is unchanging and there are no flows or external forces, then $P^\text{c}=0$, and we have the force balance for mechanical equilibrium
\begin{equation}
\Pi\s{exc} + \Pi\s{aff} + \Pi\s{el} = 0 \, .
\label{eq_polymer_force_balance_equilibrium}
\end{equation}

\noindent In this case, inside the material volume exclusion and affinity are exactly compensated by the elastic force that pulls the material inward~\cite{Katchalsky_1954}. %That we must evaluate a balance of pressures, with the pressure obtained from a differentiation of free energy with volume, is expressed in Eqs.~(8)--(13) in ref.~\cite{Katchalsky_1954} with reference to J. Frenkel (1938).

A force balance for low and moderate degrees of stretching is obtained from combination of Eqs.~\eqref{eq_bmcsl_sb},~\eqref{eq_Pi_affinity},~and~\eqref{eq_polymer_force_balance_equilibrium}, which results in
\begin{equation}
{\phi^2\left(3-\phi^2\right)}{\left(1-\phi\right)^{-3}} - \chi \phi^2 - \left(\sigma\s{b}/{k}\right) \cdot \widetilde{x}_0^2 \cdot \phi\s{0}^{2/3} \cdot \phi^{1/3}       = 0
\label{eq_force_balance}
\end{equation}
where we implemented the moderate stretching contribution to the elastic energy given by Eq.~\eqref{eq_osm_pr_hookean}. For any value of $\chi$ and of the group $\sigma\s{b}/k\cdot \widetilde{x}_0^2 \cdot \phi\s{0}^{2/3}$, Eq.~\eqref{eq_force_balance} calculates the equilibrium density of the porous material. For a very porous material (low $\phi$), we can solve Eq.~\eqref{eq_force_balance} and find that the volume of a particle \textit{V} (inversely proportional to the polymer density $\phi$) depends on Kuhn length and interaction energy $\chi$ according to
\begin{equation}
V \propto  \left\{ {k} \cdot \left(3-\chi\right) \right\}^{3/5} 
\label{eq_three_five_limit}
\end{equation}
where the $\propto$-sign refers to volume \textit{V} being proportional to the group on the right side. Eq.~\eqref{eq_three_five_limit} shows that --all other things equal-- the material swells %(volume \textit{V} goes up) 
when Kuhn length $k$ increases, and when the solvent-polymer interaction parameter $\chi$ decreases (become less positive, or more negative), i.e., when the material is more solvo-philic. Note that in this simplified model, $\chi$ cannot be larger than $\chi\!=\! 3$, but in the general model we discuss next, this limitation is not there.

In a more general model based on Eq.~\eqref{eq_polymer_force_balance_equilibrium} we no longer use the moderate stretching limit, but use the exact Langevin equation for $\widetilde{f}$. The force balance is now a set of equations that must be solved simultaneously, given by 
\begin{equation}
\left( \frac{3-\phi^2}{\left(1-\phi\right)^{3}} - \chi \right) \cdot \phi^2 - \frac{\sigma\s{b}}{k} \cdot \phi \cdot \frac{\widetilde{x}}{3} \cdot 
k\widetilde{f} = 0  \hspace{2mm}, \hspace{2mm} \widetilde{x} =\frac{1}{\tanh \left( k  \widetilde{f} \right)}-\frac{1}{k  \widetilde{f}} \hspace{2mm}, \hspace{2mm} \left(\frac{\widetilde{x}}{\widetilde{x}_0}\right)^{\text{~}3} = \frac{\phi\s{0}}{\phi}  \,  .
\label{eq_force_balance_gen}
\end{equation}

\noindent The three unknowns in these three equations are the polymer volume fraction $\phi$, the elastic force times Kuhn length, $k\widetilde{f}$, and the stretching degree $\widetilde{x}$. Input parameters in the equations are the attraction parameter $\chi$, the ratio of bead size over Kuhn length $\sigma\s{b}/k$, and the reference polymer density $\phi\s{0}$ (at an arbitrary stretching degree, $\widetilde{x}_0$). It is possible to write Eq.~\eqref{eq_force_balance_gen} as an implicit equation with $\phi$ the only unknown, after which $\widetilde{x}$ and $k\widetilde{f}$ follow directly, but that is a very unwieldy expression. Thus, Eq.~\eqref{eq_force_balance_gen} includes finite chain stretching, and an accurate equation-of-state for polymer volume exclusion.

\begin{figure} \centering
\includegraphics[width=0.7\textwidth]{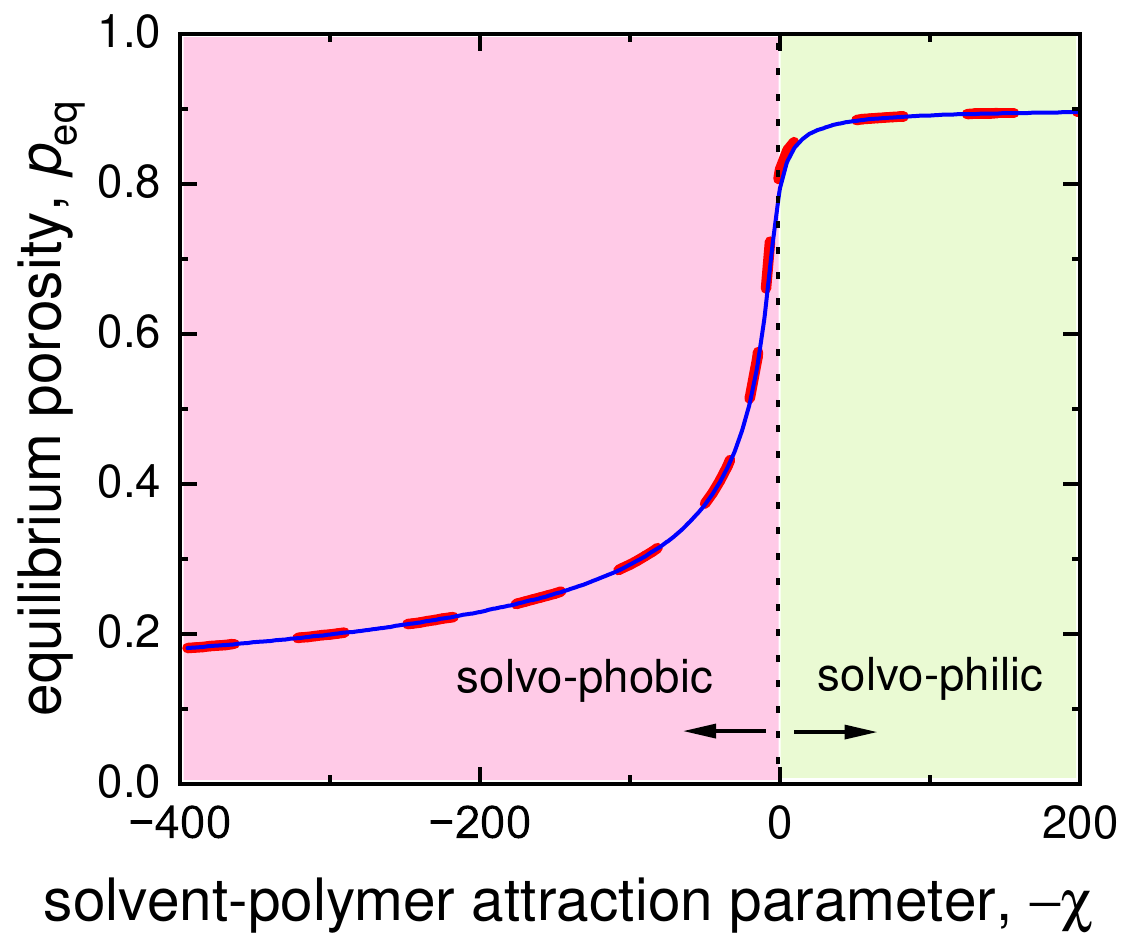}
\vspace{-10pt} 
\caption{The equilibrium porosity of a cross-linked polymer network, $p\s{eq}$, as function of the solvent-polymer attraction parameter, $\chi$, from very positive on the left (the material is very solvo-phobic) to very negative on the right (the material is very solvo-philic), in the absence of an external pressure. Parameter settings are $\phi\s{min}=\!0.10$ and $\sigma\s{b}/k\!=\!1$. The blue line is a calculation for a material free to expand in all directions, while the red dashed line is for expansion in only one direction.} 
\label{fig_1}   
%\vspace{-1em}
\end{figure}

In Fig.~\ref{fig_1} we show how the porosity, $p$, of the polymer material ($p=1-\phi$) increases when the material becomes more solvo-philic (attraction parameter $\chi$ more negative, which is to the right in the figure). On the very right we reach the maximum porosity and thus all polymer chains in the material are fully stretched, which requires that the solvent-polymer interaction is very favourable. We then reach the set value for the minimum polymer volume fraction, which in this calculation is $\phi\s{min}=0.10$, and thus the maximum porosity is $p\s{max} = 0.90$. Starting left, when the material is solvo-phobic and in a collapsed state, when we make the material less solvo-phobic and move to the right in the figure ($-\chi$ increasing), at some point the material starts to expand from a porosity around 37 vol\%, to quickly reach porosities near 90 vol\%.

Above we presented general theory to describe the expansion or compression of a particle that is unrestricted in all directions. The classical approach for this problem is Flory-Rehner (FR) theory. In the remainder of this section we rederive FR theory and show how it can be improved stepwise.

A very clear explanation of FR theory for the swelling of an hydrogel particle is presented in section 3.4 of the book by Doi and See~\cite{Doi_1996}. The balance of forces is that of elasticity and of mixing. The force due to mixing is a combination of the statistics of lattice occupation (volume exclusion), and of solvent-polymer interaction. The pressure due to mixing is given by Eq.~(2.25) in ref.~\cite{Doi_1996} (see also, for instance, Eq.~(2) in ref.~\cite{Van_der_Sman_2015})
\begin{equation}
\Pi\s{mix}= - \frac{k\s{B}T}{v\s{c}} \left[ \ln\left(1-\phi\right)+\phi+\chi\phi^2\right] \sim \frac{k\s{B}T}{v\s{c}} \left[ \left( \frac{1}{2}-\chi \right)\phi^2 +\frac{1}{3} \phi^3 + \dots \right]
\label{eq_Pi_mix_Doi}
\end{equation}
where we included $N \rightarrow \infty$, and where $v\s{c}$ is the volume of a lattice site.

\noindent For moderate stretching the elastic energy of a certain gel particle is given by Eq.~(3.70) in ref.~\cite{Doi_1996}
\begin{equation}
A\s{el,FR}= \frac{3}{2} \cdot n\s{c} \cdot k\s{B}T \cdot \left(\frac{\phi_0}{\phi}\right)^{2/3}
\label{eq_Doi_A_el}
\end{equation}
where $n\s{c}$ is the number of chains in the particle, and $\phi_0$ is the polymer volume fraction `before expansion', what seems to be an arbitrary number. In our theory $\phi_0$ is also an arbitrary density, but we do define what is the stretching degree, $\widetilde{x}_0$, at that density $\phi_0$. Indeed, when we combine Eq.~\eqref{eq_x_to_eta_min} and Eq.~\eqref{eq_Hookean_energy}, and multiply by $n\s{c}$, we obtain for our theory
\begin{equation}
A\s{el} = \frac{3}{2} \cdot n\s{c} \cdot k\s{B}T \cdot \frac{\ell}{k} \cdot \widetilde{x}_0^{\text{~}2} \cdot \left(\frac{\phi_0 }{\phi}\right)^{2/3}
\label{eq_A_el_comp_DOI}
\end{equation}
which differs from Eq.~\eqref{eq_Doi_A_el} in the terms $\widetilde{x}_0^2$ and $\ell / k$, which are absent in Eq.~\eqref{eq_Doi_A_el}. One option is to choose $\phi\s{min}$ for $\phi_0$ and then we can implement $\widetilde{x}_0 \!=\! 1$.

If we derive an elastic pressure based on Eq.~\eqref{eq_Doi_A_el}, making use of $\Pi= - \partial A\s{el} / \partial V$ (while $n\s{c}$ is taken as a constant) with $V \phi = V_0 \phi_0 $, we obtain 
\begin{equation}
\Pi\s{el,FR}= - \frac{n\s{c}}{V_0} \cdot k\s{B}T \cdot \left(\frac{\phi}{\phi_0} \right)^{1/3}
\label{eq_Pi_el_Doi}
\end{equation}
whereas Eq.~\eqref{eq_A_el_comp_DOI} leads to
\begin{equation}
\Pi\s{el} = - \frac{n\s{c}}{V\s{0}} \cdot \frac{\ell}{k} \cdot \widetilde{x}_0^{\text{~}2} \cdot {k\s{B}T}  \cdot \left(\frac{\phi}{\phi_0 }\right)^{1/3}   \, .
\label{eq_Pi_el_comp_DOI}
\end{equation}
So we have the same dependence on polymer density to the power 1/3 in Eqs.~\eqref{eq_Pi_el_Doi} and~\eqref{eq_Pi_el_comp_DOI}, but at two points there are differences in the prefactors.
 
In ref.~\cite{Doi_1996}, Eqs.~\eqref{eq_Pi_mix_Doi} and~\eqref{eq_Pi_el_Doi} are now added together and this total pressure set to zero, which results in Eq.~(3.74) in ref.~\cite{Doi_1996}, which is the classical Flory-Rehner equation
\begin{equation}
\phi + \ln\left(1-\phi\right) +  \chi \phi^2 + \nu\s{c} \cdot \left(\frac{\phi}{\phi_0}\right)^{1/3} = 0
\label{eq_mech_balance_Doi}
\end{equation}
where the prefactor is $\nu\s{c}= n\s{c}/V_0\cdot v\s{c}$, with $v\s{c}$ the volume per lattice site, see Eq.~(3.72) in~ref.~\cite{Doi_1996}. 

If we combine Eq.~\eqref{eq_Pi_mix_Doi} with the correct expression for elastic pressure, Eq.~\eqref{eq_Pi_el_comp_DOI}, we obtain an improved Flory-Rehner equation, given by
\begin{equation}
\phi + \ln\left(1-\phi\right) +  \chi \phi^2 + \nu_\text{c}^* \cdot \left(\frac{\phi}{\phi_0}\right)^{1/3} = 0
\label{eq_mech_balance_FR_improved}
\end{equation}
which is different from Eq.~\eqref{eq_mech_balance_Doi} because the new prefactor also includes a dependence on Kuhn length, \textit{k}, and chain length, $\ell$, resulting in $\nu_\text{c}^* = n\s{c} / V\s{0} \cdot \ell / k \cdot \widetilde{x}_0^{\text{~}2} \cdot v\s{c} $.

But a further improvement of FR-theory is to replace the lattice statistics by an accurate equation of state for long polymer chains, and thus use the last part of Eq.~\eqref{eq_osm_pr_hookean} for $\Pi\s{el}$, which then results in Eq.~\eqref{eq_force_balance}, and thus we arrive at
\begin{equation}
- \phi^2\cdot \frac{3-\phi^2}{\left(1-\phi\right)^3} + \chi \phi^2 +  \nu_\text{c}^{\dagger} \cdot \left( \frac{\phi}{\phi_0}  \right)^{1/3} = 0
\label{eq_mech_balance_FR_improved_again}
\end{equation}
where $\nu_\text{c}^{\dagger} =  n\s{c} / V\s{0} \cdot \ell / k \cdot \widetilde{x}_0^{\text{~}2} \cdot v\s{b} $. 

Finally, we can remove the limitation of considering only moderately stretched chains, by replacing $\Pi\s{el}$ from Eq.~\eqref{eq_Pi_el_comp_DOI} with an improved expression for the elastic pressure, for which we use here the rounded Padé (rP) approximation, Eq.~\eqref{eq_Pade_rounded}, which accurately describes the elastic force from low and moderate stretching to the infinite force at full stretching. We use the factor $\phi\s{min}$ which is the polymer density when the chains are fully stretched, and thus we can implement $\widetilde{x}_0=1$. This minimum polymer density $\phi\s{min}$ is related to $\phi_0$ according to $\phi\s{min} = \phi_0 \cdot \widetilde{x}_0^{\text{~}3}$. The volume of a material at this condition of full stretching is $V\s{max}$, and it is given by $V\s{max}\cdot \phi\s{min} = V_0 \cdot \phi_0 = V \cdot \phi$. In this case the elastic pressure becomes 
\begin{equation}
\Pi\s{el,rP} = - \frac{n\s{c}}{V\s{max}} \cdot \frac{\ell}{k} \cdot {k\s{B}T}  \cdot \left(\frac{\phi}{\phi\s{min}}\right)^{1/3} \cdot \frac{1- \nicefrac{1}{3}\cdot \left(\phi\s{min} / \phi  \right)^{2/3}}{1-\left(\phi\s{min} / \phi  \right)^{2/3}}
\label{eq_Pi_el_comp_DOI_rounded_pade}
\end{equation}
and the further improved FR-equation is now
\begin{equation}
- \phi^2\cdot \frac{3-\phi^2}{\left(1-\phi\right)^3} + \chi \phi^2 +  \nu_\text{c}^{\dagger} \cdot \left( \frac{\phi}{\phi\s{min}}  \right)^{1/3} \cdot \frac{1- \nicefrac{1}{3}\cdot \left(\phi\s{min} / \phi  \right)^{2/3}}{1-\left(\phi\s{min} / \phi  \right)^{2/3}} = 0 \, .
\label{eq_mech_balance_FR_improved_again_and_again}
\end{equation}

Making the replacement $\left({n\s{c}}/{V\s{max}}\right) \cdot \left({\ell}/{k}\right) \to \left({\phi\s{min}}/{v\s{b}}\right) \cdot \left({\sigma\s{b}}/{k}\right)$, we can rewrite Eq.~\eqref{eq_Pi_el_comp_DOI_rounded_pade} to a form that we will use further on, given by
\begin{equation}
\Pi\s{el,rP} \cdot \frac{v\s{b}}{k\s{B}T} = - \phi\s{min} \cdot \frac{\sigma\s{b}}{k}  \cdot \left(\frac{\phi}{\phi\s{min}}\right)^{1/3} \cdot \frac{1- \nicefrac{1}{3}\cdot \left(\phi\s{min} / \phi  \right)^{2/3}}{1-\left(\phi\s{min} / \phi  \right)^{2/3}} \, .
\label{eq_Pi_el_rounded_pade}
\end{equation}

\section{A model for expansion and compression in only one direction}

The previous section discussed expansion and compression of a material that is free to shrink and swell in all directions. In that situation, when a material changes its volume, all chains stretch in the same way, i.e., the stretching degree of all chains remains the same, all increasing or decreasing by the same amount. This of course assumes that all chains are of equal contour length, and that there is some swelling degree at which they are stretched the same. Then, when the material swells or contracts from that condition, all of them will remain stretched the same, more or less than before. That was the approach of the previous section, where expansion was possible in all directions.

However, when expansion is only possible in one direction, i.e., in one dimension, or `1D', the situation changes. Let us assume there is one condition (i.e., a certain polymer density) when all chains are equally stretched. We will now stretch the material in one direction (either compression or expansion) while size changes in all sideways directions are impossible, for instance because the material is very thin and tightly bound to a support structure with a large area, and it can only expand in a single direction away from the support structure. Thus the material is either laterally confined or laterally infinite~\cite{MacMinn_2016}. In this way the thickness of the thin layer changes. When the material is now stretched macroscopically in this direction away from the support layer, chains are now stretched to different degrees. This is because chains oriented exactly in the direction of expansion, expand by the same degree as the macroscopic expansion, but for instance chains that are oriented at exactly right angles, they do not stretch at all. We will consider this situation next because it applies to the behaviour of a thin porous layer, such as a membrane, tightly bound to an underlying planar support structure.

In this new problem, equations from the last section for $\Pi\s{exc}$ and $\Pi\s{aff}$ remain valid, just as all equations for the elastic force and energy of a single chain. Balances of pressure also remain valid.

We introduce the parameter $\widetilde{x}_\delta$ which is a dimensionless expansion of chains in the direction of expansion of the layer. For chains exactly oriented in the direction away from the support structure, their stretching degree $\widetilde{x}$ is equal to $\widetilde{x}_\delta$. However, chains with different orientations have a lower stretching degree; for them $\widetilde{x} < \widetilde{x}_{\delta}$. Instead of Eq.~\eqref{eq_defintion_expansion_volume}, we now have the equality
\begin{equation}
\frac{1}{\widetilde{x}_\delta}\text{d} \widetilde{x}_\delta = \frac{1}{V} \text{d} V
\label{eq_1D_1}
\end{equation}
where volume \textit{V} is the thickness of the layer, $\delta$, times the area of the layer. Making use of Eq.~\eqref{eq_1D_1}, Eq.~\eqref{eq_osm_pr_definition} becomes
\begin{equation}
\Pi\s{el} = - \frac{\partial \, V \! N \left\langle U\s{el} \right\rangle  }{\partial \, V} = - V \! N \,   \frac{\partial \left\langle U\s{el} \right\rangle }{\partial V} = - \widetilde{x}_\delta \, N \,   \frac{\partial \left\langle U\s{el} \right\rangle }{\partial \widetilde{x}_\delta }
\label{eq_1D_2}
\end{equation}
where $\left\langle U\s{el} \right\rangle$ refers to the average energy per chain.

So in the direction of expansion (away from the support structure), all chains have an expansion between 0 and $\widetilde{x}_\delta$. This distribution is flat, i.e., all values in between 0 and $\widetilde{x}_\delta$ are equally likely. We introduce a parameter $\beta$ that runs from 0 to 1, corresponding to going from 0 to $\widetilde{x}_\delta$. This parameter $\beta$ `integrates' over all chains, starting at the chains at $\beta=0$ which are at right angles to the direction of expansion, to the chains at $\beta=1$ which are exactly {in} the direction of expansion. An integration over all elastic energies of all chains, i.e., for $\beta$ from 0 to 1, leads to the average energy which we need in Eq.~\eqref{eq_1D_2}.

We already identified we need to decide on a condition, on a value of $\widetilde{x}_\delta$, where all chains are equally stretched and thus have the same $\widetilde{x}$. From that point we either shrink or expand the layer to a lower or higher $\widetilde{x}_\delta$, and then chains will have a distribution of their expansion in the direction of stretching, and also a distribution in their stretching degree, $\widetilde{x}$. To decide on this value, we assume that a layer is formed when the polymer is dissolved in a very good solvent, so we take as the condition where all chains are equally stretched, the condition of full stretching, thus at that condition for each chain $\widetilde{x}\!=\!1$, and the density of the layer is $\phi\s{min}$, and $\widetilde{x}_\delta \! = \! 1$.  

For other values of $\widetilde{x}_\delta$, the stretching degree of any chain is now given by
\begin{equation}
\widetilde{x}\left(\beta\right) = \sqrt{1 -  \beta^2 \cdot \left( 1- {\widetilde{x}_\delta}^2 \right)} 
\label{eq_1D_3}
\end{equation}
where chains described by $\beta\!=\!0$ are oriented at right angles to the macroscopic direction of expansion, and never change their stretching degree, while chains described by $\beta\!=\!1$ are in the direction of macroscopic stretching and their stretch is always given by $\widetilde{x}=\widetilde{x}_\delta$. For the starting condition that $\widetilde{x}_\delta=1$, for all $\beta$, we have $\widetilde{x} = \widetilde{x}_\delta$. But when $\widetilde{x}_\delta < 1$, $\widetilde{x}$ has values in the range from $\widetilde{x}_\delta$ to 1.

So the average elastic energy per chain is now
\begin{equation}
\left\langle U\s{el} \right\rangle = \int_0^1 U\s{el}\left(\beta\right) \text{d}\beta
\label{eq_1D_4}
\end{equation}
and we must make the differentiation $\partial \left\langle U\s{el} \right\rangle / \partial \widetilde{x}_\delta$ to calculate the last term in Eq.~\eqref{eq_1D_2} for the elastic pressure, $\Pi\s{el}$. To that end we can use
\begin{equation}
\frac{\partial \left\langle U\s{el} \right\rangle }{ \partial \widetilde{x}_\delta} =
\frac{\partial}{\partial \widetilde{x}_\delta}\left\{\int_0^1 U\s{el}\left(\beta\right) \text{d}\beta\right\}
= \int_0^1 \frac{\partial U\s{el}}{\partial \widetilde{x}_\delta}  \text{d}\beta
\end{equation}
and
\begin{equation}
\frac{\partial U\s{el}}{\partial \widetilde{x}_\delta}  = \frac{\partial U\s{el}}{\partial \widetilde{x}} \cdot  \frac{\partial \widetilde{x}}{\partial \widetilde{x}_\delta} =  \frac{\ell}{k} \cdot kf \cdot \frac{1}{\widetilde{x}} \cdot \beta^2 \cdot \widetilde{x}_\delta \, .
\end{equation}

\noindent We use the rounded Padé function, Eq.~\eqref{eq_Pade_rounded}, which results in the analytical solution
\begin{equation}
\frac{\Pi\s{el}}{k\s{B}T} = - \widetilde{x}_\delta \cdot N  \cdot  \frac{\ell}{k} \cdot  \frac{\widetilde{x}_\delta}{3} \cdot \frac{7 - \widetilde{x}_\delta^2}{1-\widetilde{x}_\delta^2} = -  \frac{\phi\s{min}}{v\s{b}}  \cdot  \frac{\sigma\s{b}}{k} \cdot  \frac{\widetilde{x}_\delta}{3} \cdot \frac{7 - \widetilde{x}_\delta^2}{1-\widetilde{x}_\delta^2}
\label{eq_rP_1D}
\end{equation}
where we also used $\widetilde{x}_\delta \cdot \phi = \phi\s{min}$. Eq.~\eqref{eq_rP_1D} can also be expressed as function of $\phi$, and then we have
\begin{equation}
\Pi\s{el} \cdot \frac{v\s{b}}{k\s{B}T} = - \phi\s{min} \cdot \frac{\sigma\s{b}}{k} \cdot \frac{\phi\s{min}}{\phi}\cdot \frac{1}{3} \cdot \frac{7 - \left(\phi\s{min}/\phi \right)^2}{1-\left(	\phi\s{min}/\phi\right)^2}
\label{eq_rP2_1D}
\end{equation}
which can be combined with the expressions for $\Pi\s{exc}$ and $\Pi\s{aff}$, and in the next section also with $P^\text{c}$. There are no additional constants compared to the model of the previous section where all chains were stretched the same. The dependence of $\Pi\s{el}$ on $\phi$, however, is different from Eq.~\eqref{eq_osm_pr_hookean}, with terms now scaling in the order $\phi^{-1}$, $\phi^{-3}$, etc.

So for one-dimensional stretching, the equation corresponding to Eq.~\eqref{eq_force_balance_gen} (which is for three-dimensional expansion) becomes
\begin{equation}
\left( \frac{3-\phi^2}{\left(1-\phi\right)^{3}} - \chi \right) \cdot \phi^3 - \phi\s{min}^2 \cdot \frac{\sigma\s{b}}{k}  \cdot \frac{1}{3} \cdot \frac{7 - \left(\phi\s{min}/\phi \right)^2}{1-\left(	\phi\s{min}/\phi\right)^2} = 0  
\label{eq_force_balance_1D}
\end{equation}
which is evaluated in Fig.~\ref{fig_1} as a red dashed line. Interestingly, the calculation result is almost identical to the output of the 3D calculation, even though the elastic part of Eq.~\eqref{eq_force_balance_1D} is very different (also when evaluated numerically) from the elastic part of~Eq.~\eqref{eq_force_balance_gen}.  

We compared the rounded Padé function for 1D stretching, Eq.~\eqref{eq_rP_1D}, with numerical calculations for 1D stretching based on the strong stretching limit, Eq.~\eqref{eq_Langevin_high_stretching_limit}, and the exact Langevin equation, Eq.~\eqref{eq_Langevin}, and the rounded Padé-function only deviates very slightly, with for instance a 2\% difference at $\widetilde{x}_\delta=0.8$. For the strong stretching function and Langevin function we did not find an analytical solution, so it is very fortunate that the analytical rounded Padé-function is so close to the exact result.

Interestingly, Taylor expansion of the rounded Padé-function of Eq.~\eqref{eq_rP_1D} around $\widetilde{x}_\delta$ results in $\Pi\s{el} \propto 7/3 \cdot \widetilde{x}_\delta + 2 \cdot \widetilde{x}_\delta^3 + \dots$, while numerical evaluation of the Langevin equation suggests that the exact Taylor expansion is $\Pi\s{el} \propto 9/4 \cdot \widetilde{x}_\delta + 9/4 \cdot \widetilde{x}_\delta^3 + \dots$.

\section{Theory for compression of a porous membrane with applied pressure}

Having set up a theory for the equilibrium swelling of a cross-linked polymer network as discussed in the previous section, we now continue to describe how this equilibrium profile changes if we apply pressure to the fluid phases outside the material. Of course nothing happens when the material, which we assume is a membrane from this point onward, is placed in a volume of liquid that is equally pressurized on all sides. Instead, we only have an effect of pressure when on one side of the membrane the liquid is pressurized to a different value than on the other side. Because the membrane is porous and we have continuous open pathways, fluid will flow across the material from the side of high hydrostatic pressure to the side of low pressure. Thus inside the membrane the hydrostatic pressure gradually decreases in a direction \textit{x}. At the same time the compression pressure, which is the pressure by which external forces push on the polymer, will go up in that same \textit{x}-direction, and thus the polymer network is increasingly compacted. We present theory to calculate by how much the material is compacted, dependent on where we are within the layer. 

After a change in applied pressure (the pressure between the two liquid phases on the two sides of the membrane), for a period of time there are changes in the polymer density profile, and fluid is flowing in or out of the material. After some time all these changes have ceased and there is no further time-dependence of the polymer density profile, $\phi\left(x\right)$, and we reach steady state. This is the situation we will consider. In steady state we have a constant volumetric solvent flux, $J$, at each position in the membrane. According to Darcy's equation, solvent flux relates to the local hydrostatic pressure gradient according to
\begin{equation}
J = - \frac{k_i}{\mu} \cdot \frac{\partial P^\text{h}}{\partial x}
\label{eq_Darcy}
\end{equation}
where $\mu$ is the viscosity of the solvent (unit Pa.s), and $x$ is the coordinate across the membrane. The permeability is $k_i$ (with unit m\textsuperscript{2}) and we use the Carman-Kozeny equation to describe the dependence of $k_i$ on porosity, \textit{p}. The Carman-Kozeny equation is based on the concept that the material consists of a large assembly of spherical particles, and is given by
\begin{equation}
k_i =  \frac{1}{180} \cdot \frac{p^3}{\left(1-p\right)^2}\cdot \sigma_\s{CK}^2 = \frac{1}{180} \cdot \left(1 +  2  p + 3 p^2  + \dots   \right) \cdot p^3 \cdot \sigma_\s{CK}^2 
\label{eq_Carman}
\end{equation}
where $\sigma_\s{CK}$ is the size of the spheres used to describe the permeability of the porous medium, conceptually comparable to the $\sigma\s{b}$ introduced in the previous section, but we do not have to set these two parameters to the same numerical value. In Ch.~8 of ref.~\cite{Biesheuvel_Dykstra_2020} it was concluded that Eq.~\eqref{eq_Carman} is accurate in the range of porosities $0\!<\!p\!<\!0.35$. A modification of Eq.~\eqref{eq_Carman} was proposed %there 
that is accurate up to $p\!=\!0.93$, by replacing one of the $\left(1\!-\!p\right)$-terms in Eq.~\eqref{eq_Carman} by $\left(1\!-\nicefrac{4}{5} {p}\right)$. However, in the present work we use the original Carman-Kozeny equation. Note that $J$ is the solvent flux, or velocity, defined per unit projected area, i.e., it is a superficial velocity. It is not an interstitial (interpore) velocity, $v\s{int}$, which --for the same $J$-- increases when porosity is lower, because \mbox{$J = p \cdot v\s{int} $}. 

The compression pressure relates to the hydrostatic pressure according to 
\begin{equation}
\frac{\partial P^\text{h}}{\partial x}+\frac{\partial P^\text{c}}{\partial x}=0
\label{overall_pressure_balance_1}
\end{equation}
which can be integrated, and then expresses that the decrease in hydrostatic pressure relative to that at the position $x\!=\!0$ (the upstream, high pressure, side of the membrane), equals the compression pressure, because at $x\!=\!0$ we have $P^\text{c}\!=\!0$. Indeed, the outside surface of the material facing the high pressure liquid phase, is not influenced by the applied pressure; the compression pressure is still zero. There is the high hydrostatic pressure but that acts on it just as much at the very outside as just within the material, so the hydrostatic pressure as such does not compact a liquid-filled material. Thus, profiles of $\phi\left(x\right)$ across a membrane will be steeper when the applied pressure increases, but on the upstream side, these profiles always start at the same value, i.e., are independent of the applied pressure.

At the upstream side of the material the material is not compacted because we have zero compression pressure, and then the calculation of the last section applies, which then results in $\left.\phi\right|_{x=0}=\phi\s{eq}$, where index `eq' refers to the mechanical equilibrium condition that applies on the upstream side of the membrane. The deeper we go into the membrane (following the coordinate axis \textit{x}), the material is increasingly compacted. This is because polymer material located between 0 and \textit{x} pushes on the polymer at position \textit{x}, and this force increases with \textit{x} because solvent flows through the polymer network and exerts a drag force on it. Thus, in the direction of solvent flow the compression pressure increases and the polymer is increasingly compressed.  

So how does knowledge of the profile of $P^\text{c}\left(x\right)$ lead to the polymer density profile, $\phi\left(x\right)$, and porosity profile, $p\left(x\right)$? To this end we solve the force balance Eq.~\eqref{eq_polymer_force_balance} at each position \textit{x} in the membrane, together with Eq.~\eqref{eq_bmcsl_sb}, Eq.~\eqref{eq_Pi_affinity}, Eq.~\eqref{eq_rP2_1D}, and  Eqs.~\eqref{eq_Darcy}--\eqref{overall_pressure_balance_1}. 

Finally we must consider that the calculation is for a given amount of polymer, so starting with a membrane at equilibrium, that has density $\phi\s{eq}$ and thickness $L\s{eq}$, this total amount is conserved when the material is compressed, and thus in the calculation we have the constraint
\begin{equation}
\int_0^L \phi  \,\text{d}x=\phi\s{eq} \, L\s{eq}
\label{eq_overall_mass_balance}
\end{equation}
where \textit{L} is the thickness of the layer during operation, when the material is being compressed.

\begin{figure} \centering
\includegraphics[width=0.6\textwidth]{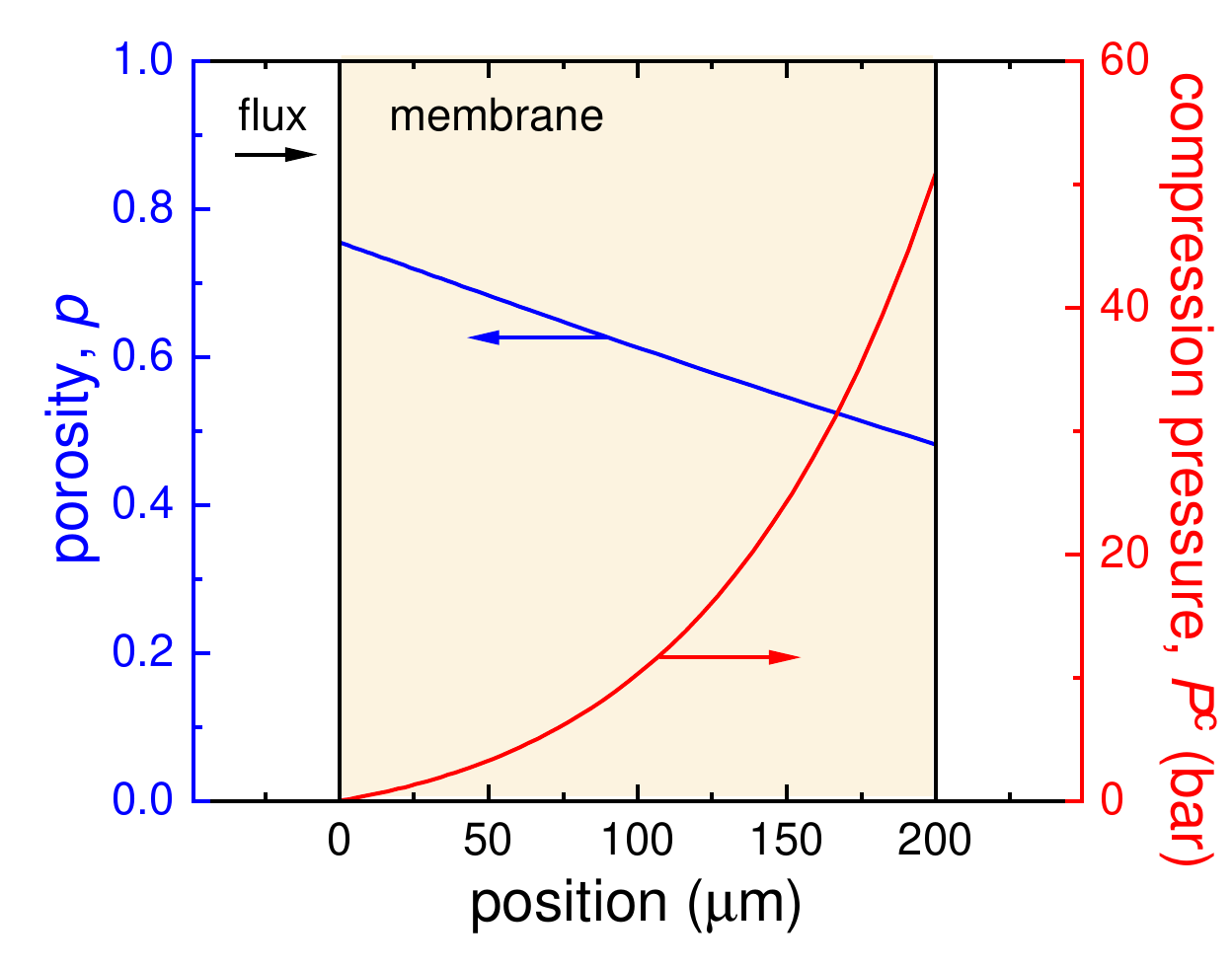}
\vspace{-10pt} 
\caption{Porosity profile across a thick membrane for conditions that lead to an overall compression by about 36\% when due to a pressure difference solvent flows across the membrane (parameter settings in main text). From the upstream to the downstream side, the porosity decreases, and the compression pressure increases, the most near the downstream side of the membrane.}
\label{fig_2}   
\end{figure}

We make a calculation for transport through a membrane with $J \! = \! 0.60$~L/m\textsuperscript{2}/hr (LMH), $\chi \!= \! 5$, $\sigma\s{b} \! = \! k \! = \! 2$~nm, $\phi\s{min}\! = \! 0.10$, $\mu\!=\! 1$~mPa.s, and $\sigma_\s{CK} \! = \! 1$~nm. In this case the equilibrium porosity is $p\s{eq}\!=\!0.754$, and thus $\phi\s{eq} \! = \! 0.246$. Calculation results are presented in Fig.~\ref{fig_2} as function of position in the membrane after the membrane is compressed to a thickness of $L \! = \! 200$~$\mu$m. Before compression it had a thickness of $L\s{eq} \! = \! 314\text{~}\mu$m. The calculation clearly shows that porosity decreases from the high pressure side to the low pressure side of the membrane, while the compression pressure strongly increases toward the downstream side of the layer to a maximum of 50 bar in this calculation.

\begin{figure} \centering
\includegraphics[width=0.6\textwidth]{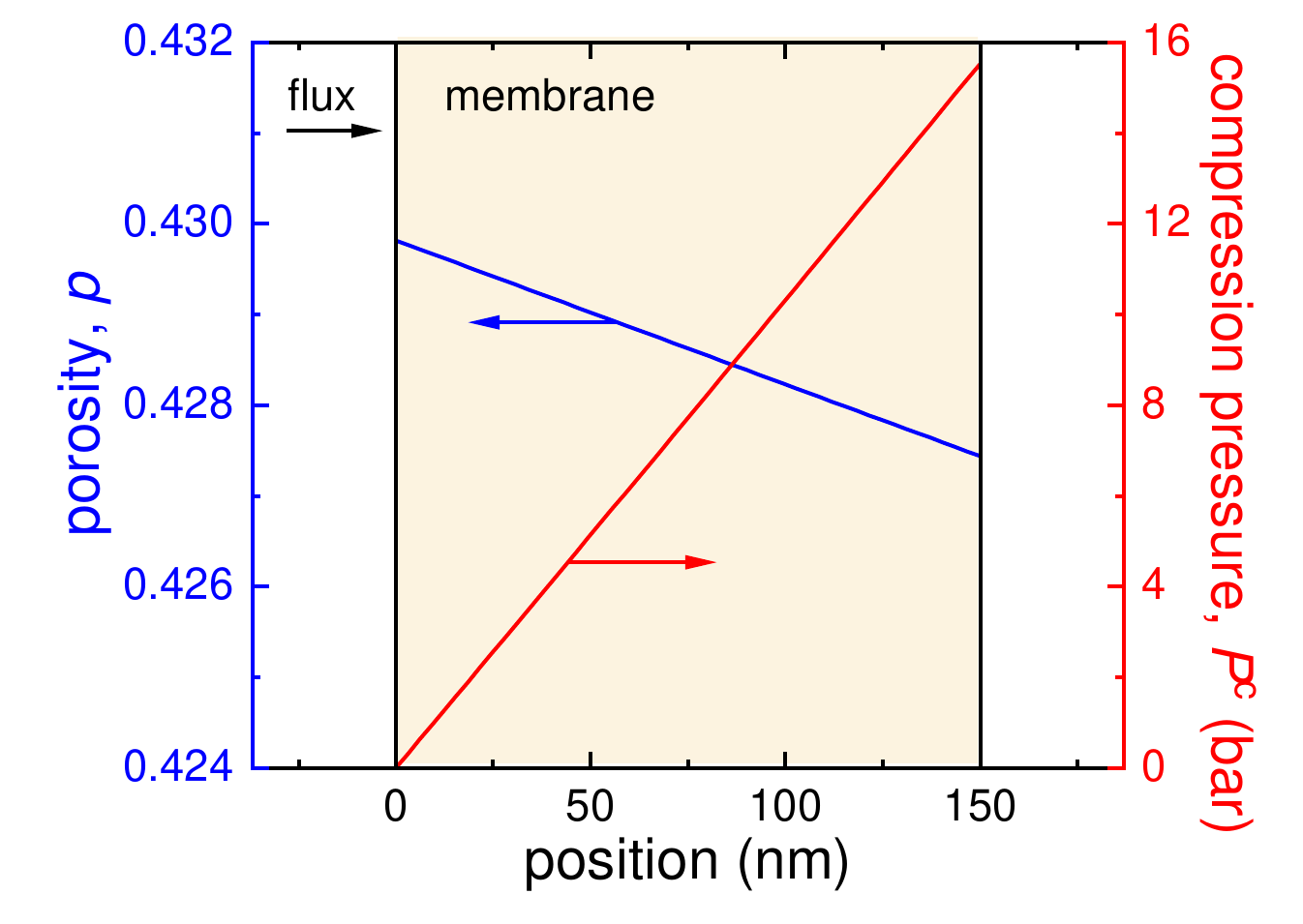}
\vspace{-10pt} 
\caption{Porosity profile across a thin membrane with water flow from left to right for parameter settings that are realistic for a brackish water reverse osmosis membrane (parameter settings in main text). The porosity only decreases very slightly across the membrane, while the compression pressure increases linearly.}
\label{fig_3}   
\end{figure}

A second calculation is for a much thinner membrane, representative of the toplayer in reverse osmosis (RO) membranes for water desalination. The result we present is for a membrane of thickness $L\s{eq}\!=\!150$~nm. Other parameter settings are $\chi \!= \! 30$ (more solvo-phobic than in Fig.~\ref{fig_2}), $\sigma\s{b} \! = \!  k \! = \! 1$~nm, $\phi\s{min}\! = \! 0.40$, $\mu\!=\! 1$~mPa.s, and $\sigma_\s{CK} \! = \! 1$~nm. The equilibrium polymer density is now $\phi\s{eq}\!=\!0.570$. The solvent permeability for this condition is $A\!=\! 3.26$~LMH/bar, a very realistic number for brackish water RO membranes. We report in Fig.~\ref{fig_3} results for a realistic transmembrane solvent flux of $J\!=\!50$~LMH. If the membrane would not be compressed, the required pressure is then $\Delta P^\text{h}\!=\! 15.35$~bar. We find in this calculation a slight decrease of porosity from~0.4300 to~0.4275. Because of this slight compression, the applied pressure is not~15.35 bar, but $\Delta P^\text{h}\!=\!15.52$~bar. These results point to the fact that the very thin membranes (selective layer) that are presently applied in the RO process for water desalination do not compress much, in line with results reported in ref.~\cite{Davenport_2020}. This conclusion is also in line with analysis of multiple data sets of water flux as function of pressure and salt concentration, that for the same membrane can always be described by one unique value of water permeability~\textit{A}~\cite{Biesheuvel_2023}. If the membrane would significantly compact at higher pressures, a good fit of these data would not be possible with a constant value of \textit{A}. Thus, also with a significant pressure-driven fluid flow through the pores of such a membrane, for the very thin RO membranes currently in use, a significant change of porosity across the thickness of the membrane is not expected.

\section{Theory for the dynamics of expansion and compression}

In this final section we analyse the dynamics of swelling or shrinking of a porous material, such as a hydrogel particle or thin layer of polymer. We make this calculation in the absence of an applied pressure difference, and instead analyse how the thickness of a layer changes when we suddenly change the solvent-polymer attraction parameter, $\chi$, from a positive value, at which porosity is low, to a negative value, when porosity is high, i.e., after the change, the material will expand. Such a change can occur when $\chi$ is very temperature-dependent and we change temperature. We use the rounded Padé equation, Eq.~\eqref{eq_Pi_el_rounded_pade} to describe isotropic swelling of a material, i.e., we assume all chains locally have the same stretching. 

So we consider the situation that a layer or particle is equilibrated at a certain condition, and everywhere has the same (low) porosity, $p_0= 1-\phi_0$. Then the temperature changes suddenly, thus $\chi$ changes suddenly, and after some time the material is fully expanded and has a high porosity, $p_\infty = 1 - \phi_\infty$. On the outside of the material, solvent can freely move in and out, and there, at that very surface, the new equilibrium density is immediately reached after the change in temperature. However, for material away from the edge it takes time to expand against the friction of the solvent that must flow in.

To analyse this problem, we can make use of most of the equations also used in the last sections, namely Eqs.~\eqref{eq_bmcsl_sb},~\eqref{eq_Pi_affinity},~\eqref{eq_polymer_force_balance},~\eqref{eq_Pi_el_rounded_pade}, and  Eqs.~\eqref{eq_Carman}--\eqref{eq_overall_mass_balance}. However, we must modify Eq.~\eqref{eq_Darcy} to~\cite{MacMinn_2016,Punter_2020}
\begin{equation}
p \cdot \left( v\s{s} - v\s{m} \right )= -  \frac{k_i}{\mu} \frac{\partial P^\text{h}}{\partial x}
\label{eq_Darcy_dyn}
\end{equation}
where $v\s{s}$ is the solvent velocity within the porous material, and $v\s{m}$ is the velocity of the porous material. Both these velocities are interstitial, i.e., a velocity `within its own phase.' There is no flow in or out of the porous material at the inside, which is at position $x\!=\!0$, so we have at all planes parallel to that boundary the overall balance of fluxes~\cite{Punter_2020}
\begin{equation}
p \cdot v\s{s} + \phi \cdot v\s{m} = 0 \, .
\label{eq_overall_continuity}
\end{equation}
When we combine these two equations with Eq.~\eqref{overall_pressure_balance_1}, we arrive at
\begin{equation}
v\s{m} = - \frac{k_i}{\mu} \cdot \frac{\partial P^\text{c}}{\partial x} \, .
\label{eq_Darcy_dyn_2}
\end{equation}
The pressure $P^\text{c}$ is given by Eq.~\eqref{eq_polymer_force_balance}, and the three contributions, $\Pi\s{exc}$, $\Pi\s{aff}$, and $\Pi\s{el}$, that must be added up to arrive at $P^\text{c}$, are given by Eqs.~\eqref{eq_bmcsl_sb},~\eqref{eq_Pi_affinity}, and~\eqref{eq_Pi_el_rounded_pade}. They are all explicit functions of polymer volume fraction $\phi$, and thus we can derive the function $f\left(\phi\right)=\partial P^\text{c} / \partial \phi$, which is lengthy but explicit in $\phi$. %, and is provided further on. 
Then Eq.~\eqref{eq_Darcy_dyn_2} becomes a function of $\phi$ and $\partial \phi / \partial \xi$, and is given by
\begin{equation}
v\s{m} = - \frac{k_i \left(\phi\right)}{\mu \cdot L_0} \cdot \frac{\partial P^\text{c}}{\partial \phi} \cdot  \frac{\partial \phi}{\partial \xi} \cdot \frac{\partial \xi}{\partial \overline{x}}
\label{eq_Darcy_dyn_3}
\end{equation}
where we write $k_i\left(\phi\right)$ to stress that also the permeability is a function of $\phi$, and we introduce the dimensionless position $\overline{x}= x/L_0$. We also introduce the material coordinate $\xi$, which runs from 0 to 1 across the material and tracks the movement of the polymer skeleton. This method is called the Lagrangian approach~\cite{MacMinn_2016}. This coordinate axis, and the associated gridpoints (or, nodes) are as it were attached to the polymer network and move with the network when it expands. Only at the very start and end of the process are they at equal distances from each other. However, at in-between times some are closer to neighbouring points and some are further apart. 

From the mass balance for a planar layer, $\phi_0  \, \xi = \int_0^{\overline{x} }\phi \, \text{d} \overline{x}$, we can derive
\begin{equation}
\frac{\partial \overline{x}}{\partial \xi} = \frac{\phi_0 }{\phi} 
\label{eq_x_xi_phi}
\end{equation}
which can replace the last term in Eq.~\eqref{eq_Darcy_dyn_3} when we consider expansion of a planar layer. After discretizing the layer in a number of gridpoints, with $i\!=\! 0$ on the inside, and $i\!=\!\text{np}$ on the outside, we can solve Eq.~\eqref{eq_Darcy_dyn_3} at all inner gridpoints to calculate the velocity of the material at these (moving) positions. At the inside, which is position $x \! = \! 0$, which is the symmetry plane in case we study a particle or stand-alone layer, or it is an impermeable surface on which the porous layer is attached, there the velocity of the polymer material is zero, and thus $\partial \phi / \partial \xi =0$ there. An overall mass balance is solved using the Trapezoid rule 
\begin{equation}
\sum_{i=1..\text{np}}  \tfrac{1}{2} \cdot \left(\phi_{i-1}+\phi_i\right)\cdot \left(\vphantom{\phi_{i-1}} \overline{x}_i - \overline{x}_{i-1}\right) = \phi_0 
\end{equation}
where again we assume that we describe a planar layer.

So we calculate the velocities in the expanding material for a range of gridpoints that span the layer from the very inside to the very outside. The velocities of these gridpoints are zero until we change the temperature, and right after that moment the points start to move because the material expands, except for the innermost point at $\overline{x} \! = \! 0$. After some time all velocities go back to zero when the material reaches the new equilibrium state. The velocity of a material gridpoint relates to its position according to
\begin{equation}
\frac{\partial \overline{x}_i }{ \partial \overline{t} }= \frac{t\s{ref} }{ L_0} \cdot v_{\text{m},i}
\end{equation}
which we solve with the Runge-Kutta or Euler backward method. The reference value for time, $t\s{ref}$, is given by
\begin{equation}
t\s{ref}= \frac{L_0^2}{\sigma_\text{CK}^2} \cdot \frac{\vphantom{L_0^2}v\s{b}}{\vphantom{\sigma_\text{CK}^2} k\s{B}T}\cdot \mu  \, .
\end{equation}
As mentioned, we solve the expansion of a planar layer, not a spherical particle. We will use a change in $\chi$ such that the layer will expand by a factor of five. For a spherical particle, going from the same initial to final $\chi$, this would only lead to a size increase of $\sim \! 70\%$. It is expected that the dynamics of size change are similar in both geometries when the initial thickness of the layer $L_0$ is chosen such that $L_0 =  R_0/3$, where $R_0$ is the initial radius of the spherical particle, because then the area/volume-ratio is the same in both geometries, at least initially. 

Analysing the equations, we note that the calculation results for the development with time $\overline{t}$ of $\overline{x}$ and $\phi$ are completely independent of the factors that are collected in $t\s{ref}$, such as initial thickness $L_0$ or $\sigma\s{CK}$. Thus, given the chosen geometry and equations, for instance those for $k_i$ and $\Pi\s{el}$, for given values of $\chi$, $\phi\s{min}$ and $\sigma\s{b}/k$, which decide on the equilibrium density, the manner in which the material expands is independent of factors such as $L_0$ or $\sigma\s{CK}$. These factor only influence via $t\s{ref}$ how fast a material goes to the new equilibrium situation. 
 
\begin{figure} \centering
\includegraphics[width=1\textwidth]{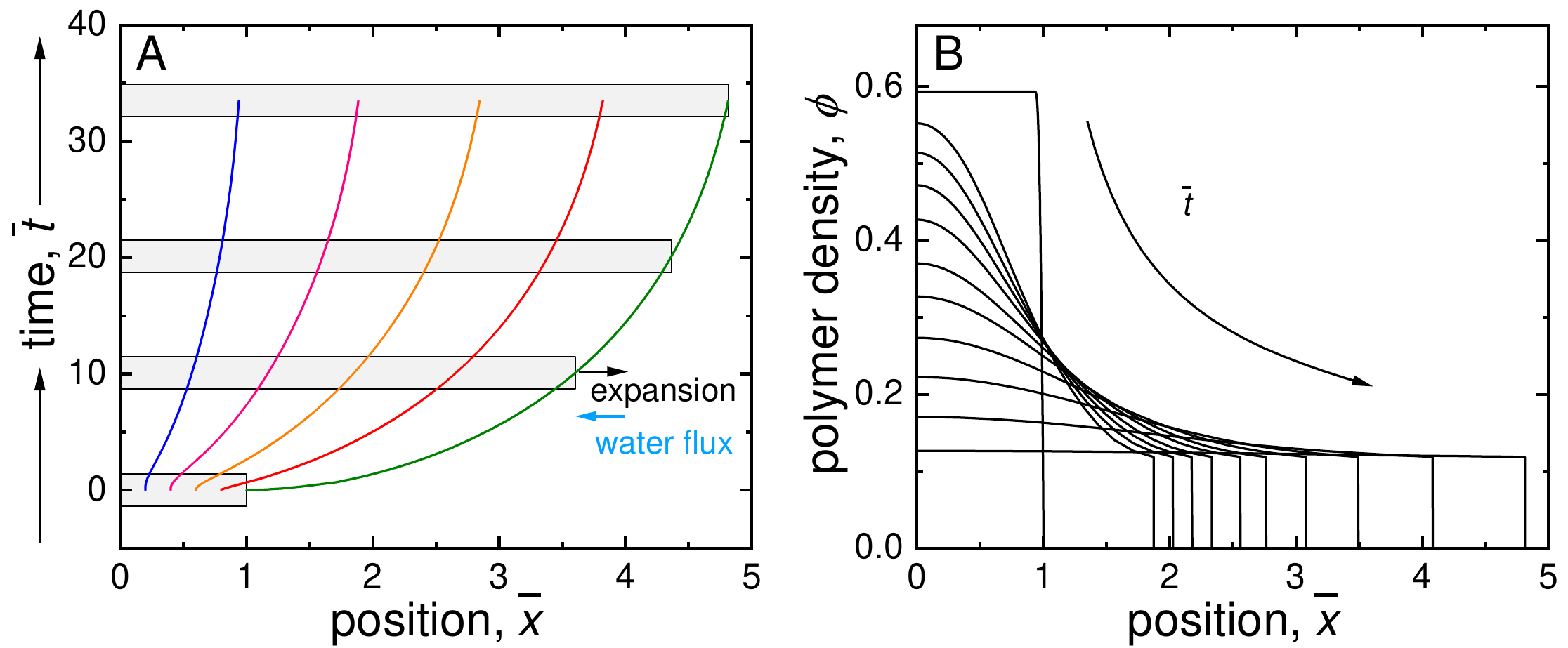}
\vspace{-10pt} 
\caption{Expansion of a planar layer of porous material upon a sudden change in polymer-solvent interaction parameter $\chi$. While the material expands, the polymer density decreases, with lower values near the outside, and higher inside. Position $\overline{x}=0$ on the left is a symmetry plane, or is where the material is placed on an impermeable surface. On the right the layer is open for free expansion and free in- and outflow of solvent. A) Trajectories of material gridpoints with time (moving upward in graph). Points of the material are tracked that are at 20, 40, 60, 80 and 100\% of the initial thickness. Initially, outer points move further apart, but eventually all points will be equidistant. B) Polymer density profiles at various times, showing decreasing profiles towards the outside, and overall a decrease in polymer density while the material expands.}
\label{fig_4}   
\end{figure}

In Fig.~\ref{fig_4}A we present calculation results according to this model, showing how in time the layer expands, first rapidly and then more slowly. Outer layers expand first, and more innermost layers follow later. As Fig.~\ref{fig_4}B shows, the polymer volume fraction, $\phi$, decreases, first in the outermost layers but soon also deeper into the layer. The calculation is based on $\phi\s{min}=0.1$ and $\sigma\s{b}/k=1$, and a value of $\chi$ chosen such that when we switch its sign at time zero, that the material will expand by a factor of five. This is the case when we change from a positive value $\chi\!=\!39.2$ before start, to a negative value $\chi\!=\!-39.2$ from time zero onward. The equilibrium polymer density then decreases from 59.4\% to 11.9\%. After the change in $\chi$, the hydrostatic pressure inside the material, $P^\text{h}$, will be less than outside, to allow water to flow inside, against the expanding porous material. The pressure at each position can be easily calculated because it is a direct function of local polymer density, $\phi$. This is because on the outside $P^\text{c}$ and $P^\text{h}$ are both zero, and $\partial P^\text{h}/\partial x + \partial P^\text{c}/\partial x=0$, so any increase in $P^\text{c}$ when moving into the layer, is equal to the decrease in $P^\text{h}$, and $P^\text{c}$ is a direct function of $\phi$. For the parameters of the present calculation, the maximum (most negative) value of $P^\text{h} \cdot v\s{b} / k\s{B}T = \overline{p}^\text{h}$ is found to be $\overline{p}^\text{h} \! = \! -27.6$. For a bead size of $\sigma\s{b}\!=\!2$~nm this translates to negative hydrostatic pressures of several 100s of bars. % and Pc 100s of bars positive. see inaugural lecture W Prins: in antquity they use peas to break rocks. 
Thus, upon a change in the polymer-solvent affinity parameter $\chi$, leading to an expansion of the material, temporarily highly negative hydrostatic pressures can develop inside a porous material. Vice-versa, if a material contracts, hydrostatic pressures inside the material can go up significantly during that process.

\section{Conclusions}

We presented various improvements of Flory-Rehner theory for the equilibrium swelling of a cross-linked porous polymer particle, and extended that theory to the description of expansion in a single direction, and finally presented theory to describe the density profile when solvent is pushed through a planar layer of such a material, i.e., a membrane. The theory includes polymer volume self-exclusion, polymer-solvent affinity, and the elastic stretching forces because of the cross-linking of the polymer network. When solvent is pushed through a membrane, a compression pressure develops, increasing in the direction of water flow, which compresses the polymer network. Because of the compression, porosity goes down and solvent-polymer friction up, and thus fluid permeability decreases. If fluid velocity is kept the same, the pressure required to push the fluid across the layer, goes up. In an example calculation for a thick membrane we show that a significant change in porosity across the membrane is possible, but for a thin membrane, the calculations show that there is hardly any change. Dynamic calculations show that when the polymer-solvent affinity of a material is suddenly increased, the porous material will expands with outer layers first adjusting to the new condition and more inner layers to follow later. An initial expansion goes very quickly but it takes very long to reach the final equilibrium state. During expansion inside a material hydrostatic pressures temporarily develop that are much lower than outside the material.

\end{document}